\documentclass[12pt,a4paper]{elsart}
\usepackage{latexsym,amssymb,epsfig}
\usepackage{amsmath,eepic}

\newcommand{\di}{\ensuremath{\mathrm{d}}}
\newcommand{\0}{$\phantom{0}$}
\newcommand{\DMS}{\mbox{\ensuremath{\mathbf{\scriptstyle^-}}\negthickspace\negthickspace {\sl D}}}
\renewcommand{\thetable}{\Roman{table}} %este sirve para la numeracion romana en las tablas

\def\0{\phantom{0}}

\begin{document}

\pagenumbering{arabic} \baselineskip25pt
\begin{center}
{\bf \large
Self-diffusion and Binary Maxwell-Stefan Diffusion Coefficients of Quadrupolar Real Fluids from Molecular Simulation} \\
\bigskip

G. A. Fern\'{a}ndez\footnotemark[1], J. Vrabec\footnotemark[1]$^,$\footnotemark[2], and
H. Hasse\footnotemark[1]
\end{center}
\footnotetext[1]{Institute of Thermodynamics and Thermal Process
Engineering, University of Stuttgart,\\ D-70550 Stuttgart,
Germany} \footnotetext[2]{To whom correspondence should be
addressed, tel.: +49-711/685-6107, fax: +49-711/685-6140, email:
vrabec@itt.uni-stuttgart.de}
\renewcommand{\thefootnote}{\alph{footnote}}

\vskip3cm Number of pages: 34

Number of tables: 3

Number of figures: 8

Running title: Diffusion coefficients from molecular simulation
\bigskip

\clearpage {\bf ABSTRACT}\\
Self- and binary Maxwell-Stefan diffusion coefficients were
determined by equilibrium molecular dynamics simulations with the
Green-Kubo method. This study covers self-diffusion coefficients
at liquid states for eight pure fluids, i.e. F$_2$, N$_2$, CO$_2$,
CS$_2$, C$_2$H$_6$, C$_2$H$_4$, C$_2$H$_2$ and SF$_6$ as well as
Maxwell-Stefan diffusion coefficients for three binary mixtures
N$_2$+CO$_2$, N$_2$+C$_2$H$_6$ and CO$_2$+C$_2$H$_6$. The fluids
were modeled by the two-center Lennard-Jones plus point-quadrupole
pair potential, with parameters taken from previous work of our
group which were determined solely on the basis of vapor-liquid
equilibrium data. Self-diffusion coefficients are predicted with a
statistical uncertainty less than 1\% and they agree within 2\% to
28\% with the experimental data. The correction of the simulation
data due to the finite size of the system increases the value of
the self-diffusion coefficient typically by 10\%. If this
correction is considered, better agreement with the experimental
data can be expected for most of the studied fluids.
Maxwell-Stefan diffusion coefficients for three binary mixtures
were also predicted, their statistical uncertainty is about 10\%. 
These results were used to test three empirical equations to estimate
Maxwell-Stefan diffusion coefficients in binary mixtures, i.e. the
equations of Caldwell and Babb, of Darken, and of Vignes.
The equations of Caldwell and Babb and of Vignes show qualitatively
different behavior of the Maxwell-Stefan diffusion coefficient
than that observed in the simulations. In agreement with previous
work, the best results are obtained in all cases with the equation
of Darken.
\\

{\bf KEYWORDS:} binary diffusion; Green-Kubo; Maxwell-Stefan; molecular dynamics;
molecular simulation; self-diffusion; quadrupole; two-center Lennard-Jones potential.
% Text body
%\section{Introduction}
\textbf{1. INTRODUCTION}

Traditionally, self-diffusion coefficients and Maxwell-Stefan
diffusion coefficients in mixtures are
obtained from empirical correlations or with more or less
theoretically based equations. Although very successful in
practical applications, this approach is limited to the range
where correlations were adjusted to experimental data and, thus, by the availability of
experimental data to fit such correlations. With increasing
computer power, molecular simulation has become an interesting
alternative tool to investigate a wide range of phenomena in many
fields of science and engineering, among which is diffusion.
The first simulation works on self-diffusion coefficients date back
to the sixties, when Alder and Wainwright \cite{alder1,alder2}
carried out simulations with hard spheres and discovered the long-time
tail of the velocity correlation function. Furthermore, Jacucci and
McDonald \cite{jaccuci}, Jolly and Bearman, and Schoen and
Hoheisel \cite{jolly,schoen} carried out computations of the
binary transport coefficients, and investigated the contribution
of the cross correlation to the binary Maxell-Stefan (MS)
diffusion coefficient. These works established the calculation
methodology and paved the way for posterior works aimed to predict
diffusion coefficients. More recently and from an engineering
point of view, Stoker and Rowley \cite{stoker,rowley1} used
molecular simulation to calculate binary MS diffusion coefficients
of binary alkane  mixtures. They proposed calculating binary MS
diffusion coefficients from self-diffusion coefficient data.

In recent work of our group, it was shown that the Lennard-Jones (LJ) potential,
adjusted only to experimental vapor-liquid equilibria, satisfactorily predicts
the self- and binary MS diffusion coefficients
\cite{fernandez1}, shear viscosities, and thermal conductivities
\cite{fernandez2} of several simple fluids and their mixtures.
These results confirm the known suitability of the spherical LJ potential to
describe these fluids \cite{mcdonald}, and also show that the
determination of the potential parameters from vapor-liquid equilibria is an
adequate choice to predict transport properties with reasonable
accuracy, at least for simple fluids.

Here, this investigation is extended to more complex molecules.
The intermolecular interactions are described by the two-center Lennard-Jones plus
point-quadrupole (2CLJQ) potential. This model has been employed
successfully by several authors, for modeling thermodynamic
properties and the self-diffusion coefficients of simple real
fluids \cite{cheung,vogelsang,hoheisel,moeller,vrabec}. Albeit the
2CLJQ potential is not new, the prediction of transport
properties with such a model has still not been explored in
detail. In order to investigate the suitability and performance of
the 2CLJQ potential with respect to self-diffusion coefficients, they were
calculated in the present work for a range of molecular fluids
(F$_2$, N$_2$, CO$_2$, CS$_2$, C$_2$H$_6$, C$_2$H$_4$, C$_2$H$_2$,
SF$_6$) and compared to existing experimental data for these
fluids. Good predictions of the self-diffusion coefficients were
observed in most cases. Also self- and MS diffusion coefficients
for the binary mixtures N$_2$+CO$_2$, N$_2$+C$_2$H$_6$ and
CO$_2$+C$_2$H$_6$ were studied. These results were used to
evaluate the performance of three equations for describing binary
MS diffusion coefficients, namely the equations of Caldwell and
Babb \cite{caldwellbabb}, Darken \cite{darken}, and Vignes \cite{vignes}. A direct
comparison of simulation results to experimental data of binary MS diffusion coefficients
is not possible for the fluids studied here, because of the lack of
such data.

\clearpage

%\section{Method}
%\subsection{Molecular models}
\textbf{2. METHOD}\\
2.1. Molecular Model

In the present work, interactions between molecules are described
by two-center Lennard-Jones plus point-quadrupole (2CLJQ) based
potential models. These models have recently been
developed in our group \cite{vrabec} as part of a study covering
25 pure substances. The 2CLJQ model is a pairwise additive
potential model consisting of two Lennard-Jones sites a distance
$L$ apart plus a point-quadrupole of moment $Q$ located in the
geometric center of the molecule and oriented along the molecular axis,
which connects the two LJ sites. The interaction
energy of two molecules $i$ and $j$ is

\begin{equation}\label{tpoten}
u^{\rm 2CLJQ}_{ij} = \sum_{a=1}^{2} \sum_{b=1}^{2} 4\epsilon_{ij}
\left[ \left( \frac{\sigma_{ij}}{r_{ab}} \right)^{12} - \left(
\frac{\sigma_{ij}}{r_{ab}} \right)^6 \right] + u_{\rm
Q}(r_{ij}, \theta_i, \theta_j, \phi_{ij}, Q).
\end{equation}

Here, $r_{ab}$ is the distance between LJ site $a$ and LJ site $b$; $a$
counts the two sites of molecule $i$, $b$ counts those of molecule
$j$. The LJ parameters $\sigma_{ij}$ and $\epsilon_{ij}$ represent the size and
energy parameters of the LJ potential, respectively. The quadrupolar contribution is given by
\cite{graygubb}

\begin{eqnarray}\label{qpoten}
u_{\rm Q}(r_{ij}, \theta_i, \theta_j, \phi_{ij}, Q)=\frac{3}{4}
\frac{Q^2}{r_{ij}^5} \left[ 1-5 \left(
c_i^2+c_j^2 \right) -15 c_i^2 c_j^2 + 2 \left( s_i s_j c - 4 c_i
c_j \right)^2 \right],
\end{eqnarray}

with $c_k={\rm cos} \theta_k$, $s_k={\rm sin} \theta_k$, and
$c={\rm cos} \phi_{ij}$. Here, $r_{ij}$ is the
center-center distance of the two molecules $i$ and $j$.
$\theta_i$ is the angle between the axis of the molecule $i$ and the center-center
connection line, and $\phi_{ij}$ is the azimutal angle between the
axis of molecules $i$ and $j$. More details can be found in Gray and
Gubbins \cite{graygubb}.

Pure substance parameters $\sigma_{ii}$ and $\epsilon_{ii}$ were
taken from Ref. 15 and are summarized in Table \ref{T1}. They were
adjusted to experimental vapor pressure and saturated liquid density
data of the pure substance. For symmetric diatomic molecules fluorine (F$_2$) and
nitrogen (N$_2$), and symmetric triatomic molecules like carbon
dioxide (CO$_2$) and carbon disulfide (CS$_2$), as well as (C$_2$)
derivates as ethane (C$_2$H$_6$) and ethylene (C$_2$H$_4$) the
description of the interaction by the 2CLJQ represents a good approximation.
However, since SF$_6$ molecules are neither elongated nor quadrupolar, the
fitted parameters obtained for the 2CLJQ model lose all physical meaning.

For the modeling of mixtures, the like interactions are fully described by the
pure substance models. The same holds for the unlike quadrupolar interaction, which
is exactly determined by electrostatics, cf. Eq. (\ref{qpoten}). On the other hand,
the parameters of the unlike LJ interactions are obtained from the pure fluid parameters
by the modified Lorentz-Berthelot combination rule

\begin{equation}\label{berte}
\sigma_{12}=\frac{(\sigma_{11}+\sigma_{22})}{2}
\end{equation}
and
\begin{equation}\label{lorenz}
\epsilon_{12}=\xi \cdot \sqrt{\epsilon_{11}\epsilon_{22}},
\end{equation}

where $\xi$ is a binary interaction parameter that was adjusted to
one experimental bubble point of the binary mixture. It has been shown in previous
work of our group for numerous systems \cite{vrabec2,vrabec3,stoll2} that binary and ternary vapor-liquid
equilibria can be described accurately in this way.
The parameters used in this work were taken from Ref. 22 and their values are
1.041, 0.974, 0.954 for N$_2$+CO$_2$, N$_2$+C$_2$H$_6$, and CO$_2$+C$_2$H$_6$, respectively.
\clearpage

%\subsection{Diffusion coefficients}
2.2. Diffusion Coefficients

Diffusion coefficients can be calculated by equilibrium molecular
dynamics with the Green-Kubo formalism \cite{green,kubo}. In this
formalism, transport coefficients are related to integrals of
time-correlation functions of the corresponding fluxes. There are various methods to relate
transport coefficients to time-correlation functions; a good
review was given by Zwanzig \cite{zwanzig}. The self-diffusion
coefficient of a molecular fluid is characterized by the mass
current of a single target molecule \cite{hoheisel1}. It is given
by

\begin{equation}
\label{self} D_i=\frac{1}{3N_i}\int_{0}^{\infty} \di t
\Big\langle\sum_{k=1}^{N_i} \mathbf{v}_i^k(0) \cdot
\mathbf{v}_i^k(t) \Big\rangle,
\end{equation}

where $\mathbf{v}_i^k(t)$ expresses the velocity vector of the center of mass
of molecule $k$ of species $i$, and $<...>$ denotes an ensemble average. Equation (\ref{self}) yields the
self-diffusion coefficient for component $i$ by averaging over $N_i$ molecules. Also, the
expression for the binary MS diffusion coefficient $\DMS_{12}$ is given in terms of
velocities of the molecular the centers of mass

\begin{equation}
\label{mutual} \DMS_{12} = \frac{x_2}{3N_1} \left(\frac{x_1 M_1+x_2 M_2}{x_2M_2}
\right)^2 \int_{0}^{\infty} \di t~\Big\langle \sum_{k=1}^{N_1} \mathbf{v}_1^k(0) \cdot
\sum_{k=1}^{N_1} \mathbf{v}_1^k(t) \Big\rangle,
\end{equation}

where $M_i$ denotes the molar mass and $x_i$ the mole
fraction of species $i$.\\
The present simulations yield both self-diffusion coefficients and
binary MS diffusion coefficients. Unfortunately, a direct comparison
between the simulated and experimental binary MS diffusion
coefficients is not possible for the investigated mixtures due to
the absence of experimental data. Nevertheless, it is possible to
estimate the binary MS diffusion coefficients from empirical
equations that relate the self-diffusion coefficients or infinite
dilution binary diffusion coefficients to the binary MS diffusion coefficients
through simple functions of the composition. Here, three such
equations are considered: Darken's equation \cite{darken},
Caldwell and Babb's equation \cite{caldwellbabb}, and Vignes'
equation \cite{vignes}. Darken's equation relates the
self-diffusion coefficients of both components $D_{1}$ and $D_{2}$
to the binary MS diffusion coefficient $\DMS_{12}$

\begin{equation}
\label{darken} \DMS_{12}= D_{1} \cdot x_2+D_{2} \cdot x_1.
\end{equation}
It is important to note that the self-diffusion coefficients are
needed for each studied composition so that Eq. (\ref{darken}) is
only of limited use for practical aplications. Vignes' equation
\cite{vignes} and Caldwell and Babb's equation \cite{caldwellbabb}
relate the MS diffusion coefficients to the infinite dilution binary diffusion
coefficients $D_{12}^\infty$ and $D_{21}^\infty$. The Caldwell and
Babb equation is given by

\begin{equation}
\label{cababb} \DMS_{12}= D_{21}^\infty \cdot {x_1}+ D_{12}^\infty
\cdot x_2,
\end{equation}

and the Vignes equation by

\begin{equation}
\label{vignes} \DMS_{12}= (D_{21}^\infty)^{x_1} \cdot
(D_{12}^\infty)^{x_2}.
\end{equation}

Here $D_{ij}^\infty$ is the diffusion coefficient of species $i$
infinitely diluted in species $j$. In contrast to Darken's
equation, the equations of Caldwell and Babb and of Vignes need
only two values for the whole range of composition, which makes
them attractive for practical applications.
In the limit of infinite dilution, the binary MS diffusion coefficient
and the self-diffusion coefficient coincide. This result can be obtained from Eq. (\ref{darken}),
by taking the limit $x_i\rightarrow0$, i.e. if $x_1\rightarrow0$ then
$\DMS_{12}= D_{12}^\infty=D_1$, or if $x_2\rightarrow0$ then $\DMS_{21}= D_{21}^\infty=D_2$.
This equivalence is used to obtain the self-diffusion coefficients in the infinite dilution limit.

\clearpage
%\subsection{Simulation details}
2.3. Simulation Details

Molecular simulations were performed in a cubic box of volume $V$
containing $N=500$ molecules whose interactions are described by the 2CLJQ potential. The
cut-off radius was set to $r_{c}=5 \sigma $ and the molecules
were assumed to have no preferential relative orientations outside
the cut-off sphere. For the calculation of the LJ long range
corrections, orientational averaging was applied with equally
weighted relative orientations as proposed by Lustig
\cite{lustig}. The assumption of no preferential relative
orientations beyond the cut-off sphere implies for the
quadrupolar interactions that long range corrections are not needed
since they vanish. The simulations were started from a
face-centered-cubic lattice configuration with randomly distributed velocities,
the total momentum of the system was set to zero, and modified Newton's
equations of motion were solved with the Gear predictor-corrector
integration scheme of fifth order \cite{haile}. The time step for
this algorithm was set to $\Delta t \cdot \sqrt{\epsilon_1 /
m_1}/\sigma_1=0.001$. The time-correlation functions were calculated
in the $NVT$ ensemble using the Nos\'{e}-Hoover thermostat
\cite{nose,frenkel} with a thermal inertial parameter of 10 kJ
mol$^{-1}$ ps$^2$, and the diffusion coefficients were then obtained by using Eqs. (\ref{self}) and (\ref{mutual}).
It must be pointed out that both $NVE$ and $NVT$ simulations were
performed, and the obtained diffusion coefficients agreed in all
cases within their uncertainties. It was concluded that
the Nos\'{e}-Hoover thermostat does not influence the values of the diffusion
coefficients. As $NVT$ simulations yield diffusion coefficients
exactly at the desired temperature, they were preferred. The
simulations were equilibrated in a $NVT$ ensemble over
$100~000$ to $150~000$ time steps. Once equilibrium has been
reached, the self-diffusion and MS diffusion coefficients were
evaluated. To calculate the binary MS diffusion coefficients at
the desired $T$ and $p$, a prior $NpT$ simulation \cite{andersen}
was performed, from which the density for the $NVT$
ensemble was taken. The statistical uncertainty of the diffusion
coefficients was estimated using the method of Fincham et al. \cite{fincham}.
In order to calculate the self-diffusion and
binary diffusion coefficients, similar criteria as in Ref. 8 were
applied. Self-diffusion coefficients were calculated by averaging
over $100~000$ independent autocorrelation functions, i.e. over $200$ time
origins. The time origins
were taken every 500th time step during the period of production. Depending
on the density, this distance between time origins was extended in order to
ensure their independence. The correlation function was calculated over 2500 time steps
in order to minimize the error due to the long-time tail. From pilot runs with
different lengths of correlations functions, i.e. 1000, 2500, and 3500 time steps, this error
was estimated to be about 3\%. For the calculation of the binary MS diffusion coefficients,  $12~000$
independent time origins were averaged, here a compromise
between accuracy and simulation time was made. The time origins were
taken every 100th time step, and the correlation function was calculated over 1000 to 1500
time steps. This requires simulations of about $1 \cdot 10^5$ to $4 \cdot 10^5$
time steps for the self-diffusion coefficients and $12 \cdot 10^5$ for MS diffusion coefficients.
The binary MS diffusion coefficients were calculated
for mole fractions between 0.1 and 0.9. To obtain the binary MS
diffusion coefficients at infinite dilution, a polynomial function was fitted to the
simulation results between mole fractions 0.1 and 0.9 and then
extrapolated to zero and one, respectively. The relative error
was estimated as being the same as for the
binary MS diffusion coefficients at 0.1 and 0.9, respectively.

An important issue is the influence of the moments of
inertia of the molecules on the self-diffusion coefficient. In all cases, the
experimental molecular mass \cite{nist} was distributed equally between the
two LJ centers. However, for CO$_2$ this matter was investigated. For CO$_2$,
the experimental molecular mass was distributed between the two LJ centers and the
quadrupolar site, so that the mass of the two oxygen atoms was distributed
between the two LJ centers, and the mass of the carbon atom was associated to the
quadrupolar site. In this case the tensor of moments of inertia in a
reference system with origin in the geometrical center of the
CO$_2$ molecule is diagonal, whose two nonzero elements are given
by 4.000 $L^2$ g$~$mol$^{-1}$ m$^2$. On the other hand, if the moment of
inertia is calculated sharing the total molecular mass between the
two LJ centers only, the diagonal elements have a value of 5.501 $L^2$ g$~$mol$^{-1}$
m$^2$. No difference for the self-diffusion coefficients was
found for the two different choices. This
result is plausible, because the self-diffusion coefficient is
related to the translational motion of the molecular center of
mass. \clearpage

%\section{Results}
%\subsection{Self diffusion coefficients}
\textbf{3. RESULTS}\\

In this section, the predictions for self- and binary MS diffusion
coefficients are compared to experimental data and to the
empirical equation of Liu et al. \cite{liu}, which is a
correlation based on molecular simulation results and experimental
data. The results are presented in terms of the product of
self-diffusion coefficient and density rather than the self-diffusion coefficient itself, because
the latter tends to infinity in the zero density limit. The self-diffusion
coefficient is a single-particle property, thus highly accurate data can be obtained with modest
computing time. The uncertainty of the present self-diffusion data is lower than 1\%, numerical
values for all fluids are given in Table \ref{T2}.

3.1. Self Diffusion Coefficients in Pure Fluids

\begin{figure}[ht]
\caption[Self-diffusion coefficients of fluorine, nitrogen, carbon dioxide, and carbon disulfide as predicted by
molecular simulation in comparison to experimental data \cite{oreilly,krynicki,etesse,woolf}. F$_2$ (saturated
liquid, $T$=54-105 K): ${\square}$ exp., ${\blacksquare}$ sim.; N$_2$ (saturated liquid, $T$=77-93 K): ${\circ}$
exp., ${\bullet}$ sim.; CO$_2$ (homogeneous liquid, $T$= 273 K): ${\vartriangle}$ exp., ${\blacktriangle}$ sim.;
CS$_2$ (homogeneous liquid, $T$=298.2 K): ${\triangledown}$ exp., ${\blacktriangledown}$ sim. The lines represent
the correlation of Liu et al. \cite{liu}.]{} \label{fig1}
\begin{center}
\includegraphics[width=150mm,height=200mm]{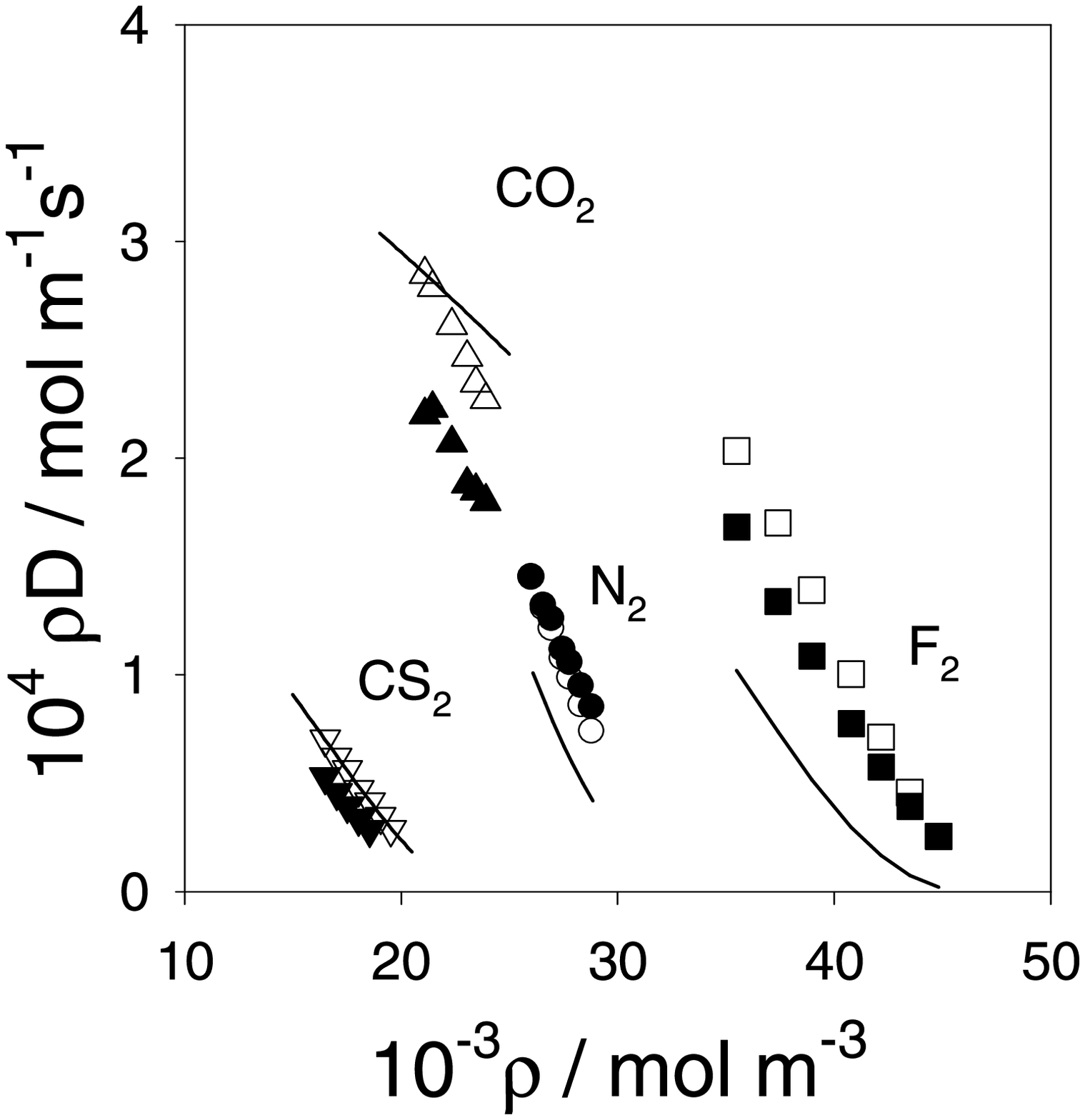}
\end{center}
\end{figure}

Figure \ref{fig1} shows the results for the product of density and self-diffusion
coefficient of F$_2$, N$_2$, CO$_2$, and CS$_2$ compared to
experimental data \cite{oreilly,krynicki,etesse,woolf}. For F$_2$
and N$_2$, the considered state points correspond to the saturated liquid, for
which experimental densities were taken from Refs. 39 and 40.
%\cite{vargaftik,refprop}
For CO$_2$ and CS$_2$, the state points
lie in the homogeneous liquid region at temperatures of 273 K
and 298.2 K, respectively. Overall, fair agreement between
experimental data and the predictions by molecular simulation is found.
The best results are obtained for N$_2$ with an average
deviation of only 6\%. For F$_2$, the predictions match the
experimental data at high densities, at low densities deviations up to 20\%
occur. The predictions for CO$_2$ are too low by about 20\%.
For CS$_2$, the predictions are also too low by about the same
amount, in this case the correlation of Liu shows better
agreement with the experimental data. It should be noted that the
poorer performance of the CO$_2$ and CS$_2$ models is reasonable since the three atoms of
roughly the same size have not been explicitly considered by the
2CLJQ model.

\begin{figure}[ht]
\caption[Self-diffusion coefficients of ethane, ethylene,
acetylene, and sulfur hexafluoride as predicted by molecular
simulation in comparison to experimental data
\cite{woolf,greiner,arends,scheie}. C$_2$H$_6$ (homogeneous
liquid, $T$=273 K): ${\square}$ exp., ${\blacksquare}$ sim.;
C$_2$H$_4$ (homogeneous liquid, $T$=298.15 K): ${\circ}$ exp.,
${\bullet}$ sim.; C$_2$H$_2$ (saturated liquid, $T$= 192-222 K):
${\vartriangle}$ exp., ${\blacktriangle}$ sim.; SF$_6$ (saturated
liquid, $T$=240-310 K): ${\triangledown}$ exp.,
${\blacktriangledown}$ sim. The lines represent
the correlation of Liu et al. \cite{liu}.]{} \label{fig2}
\begin{center}
\includegraphics[width=150mm,height=200mm]{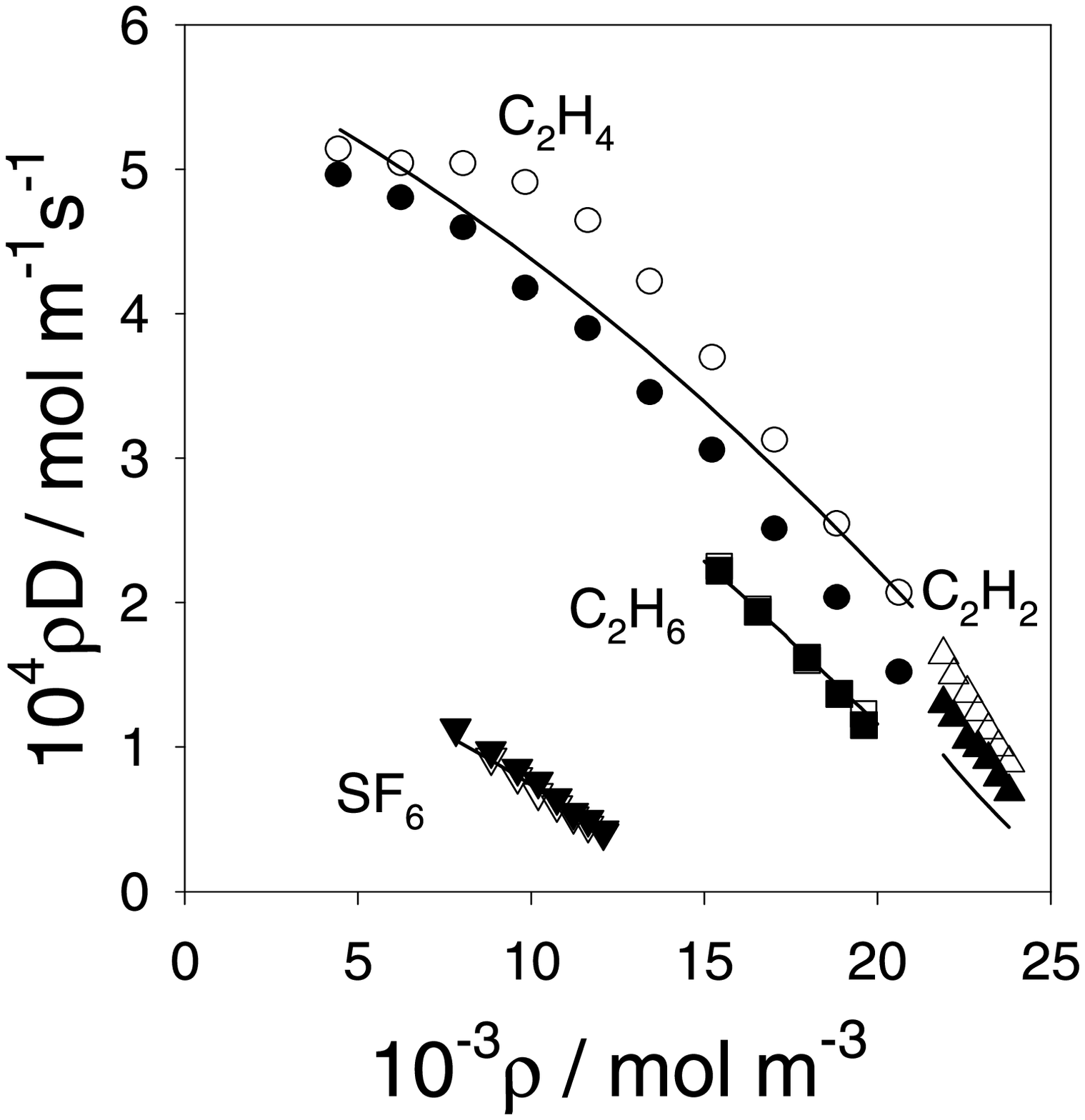}
\end{center}
\end{figure}

Figure \ref{fig2} shows the results for the product of density and self-diffusion
coefficient of C$_2$H$_6$, C$_2$H$_4$, C$_2$H$_2$, and SF$_6$
compared to experimental data \cite{greiner,arends,scheie,tison}
and Liu's correlation. For C$_2$H$_6$ and C$_2$H$_4$ the considered
state points lie in the homogeneous liquid region at temperatures of 273 K
and 298.15 K, respectively. For C$_2$H$_2$ and SF$_6$ the states
correspond to the saturated liquid, the C$_2$H$_2$ densities
were taken from Ref. 45. %\cite{daubert}
Good agreement with the
experimental data is found. The best results are found for
C$_2$H$_6$ and SF$_6$ with average deviations of only 2\% and 6\%.
For C$_2$H$_2$, the predictions of the simulation are too low
by about 20\%, for C$_2$H$_4$ they are also too low by about 15\%.
The experimental data of C$_2$H$_4$ show a pronounced
curvature that is neither reproduced by the simulations nor by
Liu's correlation. Liu's correlation is as good as the simulation for SF$_6$ and
C$_2$H$_6$, worse than the simulation for C$_2$H$_2$, but
slightly better for C$_2$H$_4$.

To study the dependence of the self-diffusion coefficient on the number of particles, one state
point for N$_2$ at $T$=85 K, $\rho$=27.526 10$^3$ mol $\cdot$ m$^{-3}$ was chosen. For this state point,
a sequence of simulations with increasing number of particles: $N$=108, 256, 500, 864, and 1372
was carried out. The values for the self-diffusion coefficients were
3.78(6), 3.96(5), 4.03(1), 4.13(2), 4.21(2) in 10$^{9}$m$^2 \cdot$ s$^{-1}$, respectively. An
estimate of the self-diffusion coefficient for an infinite system size can be obtained by a
linear fit of the self-diffusion coefficient data as a function of the inverse box
length \cite{inchul}. This fit yields a value of 4.50(4) 10$^{-9}$ m$^2 \cdot$ s$^{-1}$
for an infinitely large system, that is about 10\% larger than the results with $N$=500 particles.
As most predictions of self-diffusion coefficients are below the experimental data,
the finite-size correction can improve the agreement with the experimental data for most fluids.
Exceptions are F$_2$, SF$_6$, and C$_2$H$_6$, for which the deviations would increase.
\clearpage

%\subsection{Binary Maxwell-Stefan Diffusion Coefficients}
3.2. Binary Maxwell-Stefan Diffusion Coefficients

In this section, the results obtained for the binary mixtures
N$_2$+CO$_2$, N$_2$+C$_2$H$_6$, and CO$_2$+C$_2$H$_6$ at 253.15 K
and 20 MPa are presented. Numerical data are given in Table \ref{T3}, self-diffusion
coefficients of pure fluids in binary mixtures are reported with statistical
uncertainties less than 1\%; binary MS diffusion coefficients are reported 
with statistical uncertainties of about
10\%. These mixtures were selected since their vapor-liquid equilibria
were successfully calculated with the
present molecular models \cite{stoll2}. The simulated MS diffusion
coefficients are compared with the predictions from the equations
of Darken, Caldwell and Babb, and Vignes, cf. Eqs. (\ref{darken}),
(\ref{cababb}), and (\ref{vignes}). To evaluate their performance,
the average relative deviation,
$\sum_i(\DMS_{12,i}^{sim}-\DMS_{12,i}^{equation})/\DMS_{12,i}^{sim}$,
was calculated. Experimental data for comparison are unfortunately
not available. The input needed for Eqs. (\ref{darken}) to
(\ref{vignes}) were therefore simulation data, i.e. self-diffusion
coefficients for Darken's equation and infinite dilution diffusion
coefficients for the equations of Caldwell and Babb and of Vignes.

\begin{figure}[ht]
\caption[Binary MS diffusion coefficients for the mixture
N$_2$+CO$_2$ at 253.15 K and 20 MPa as predicted by molecular
simulation in comparison to empirical equations.]{} \label{fig3}
\begin{center}
\includegraphics[width=150mm,height=200mm]{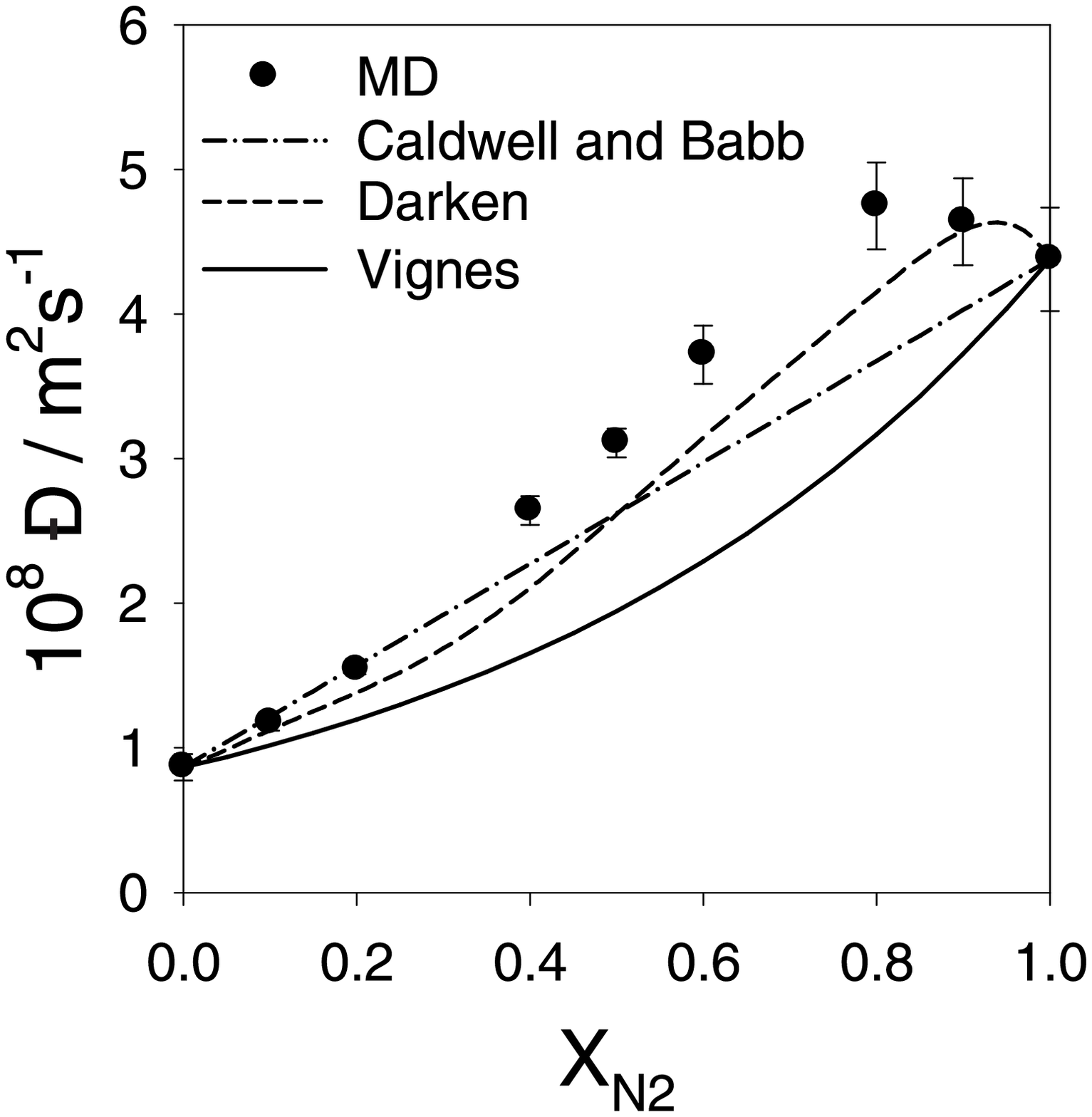}
\end{center}
\end{figure}

Figure \ref{fig3} shows the results for the binary MS diffusion
coefficients for the mixture N$_2$+CO$_2$ compared to the
equations of Caldwell and Babb, Darken, and Vignes. The MS
diffusion coefficient increases as the mole fraction of N$_2$
increases due to the smaller size and mass of the N$_2$ molecule.
The simulation results lie above the linear interpolation between
the infinite dilution diffusion coefficients, i.e. Caldwell and Babb's
equation. Vignes' equation gives a different behavior, with
negative deviations from the linear interpolation, whereas
Darken's equation predicts positive deviations from the linear
interpolation for high N$_2$ mole fractions and negative
deviations for low mole fractions.

\begin{figure}[ht]
\caption[Binary MS diffusion coefficients for the mixture
N$_2$+C$_2$H$_6$ at 253.15 K and 20 MPa as predicted by molecular
simulation in comparison to empirical equations.]{}
\label{fig4}
\begin{center}
\includegraphics[width=150mm,height=200mm]{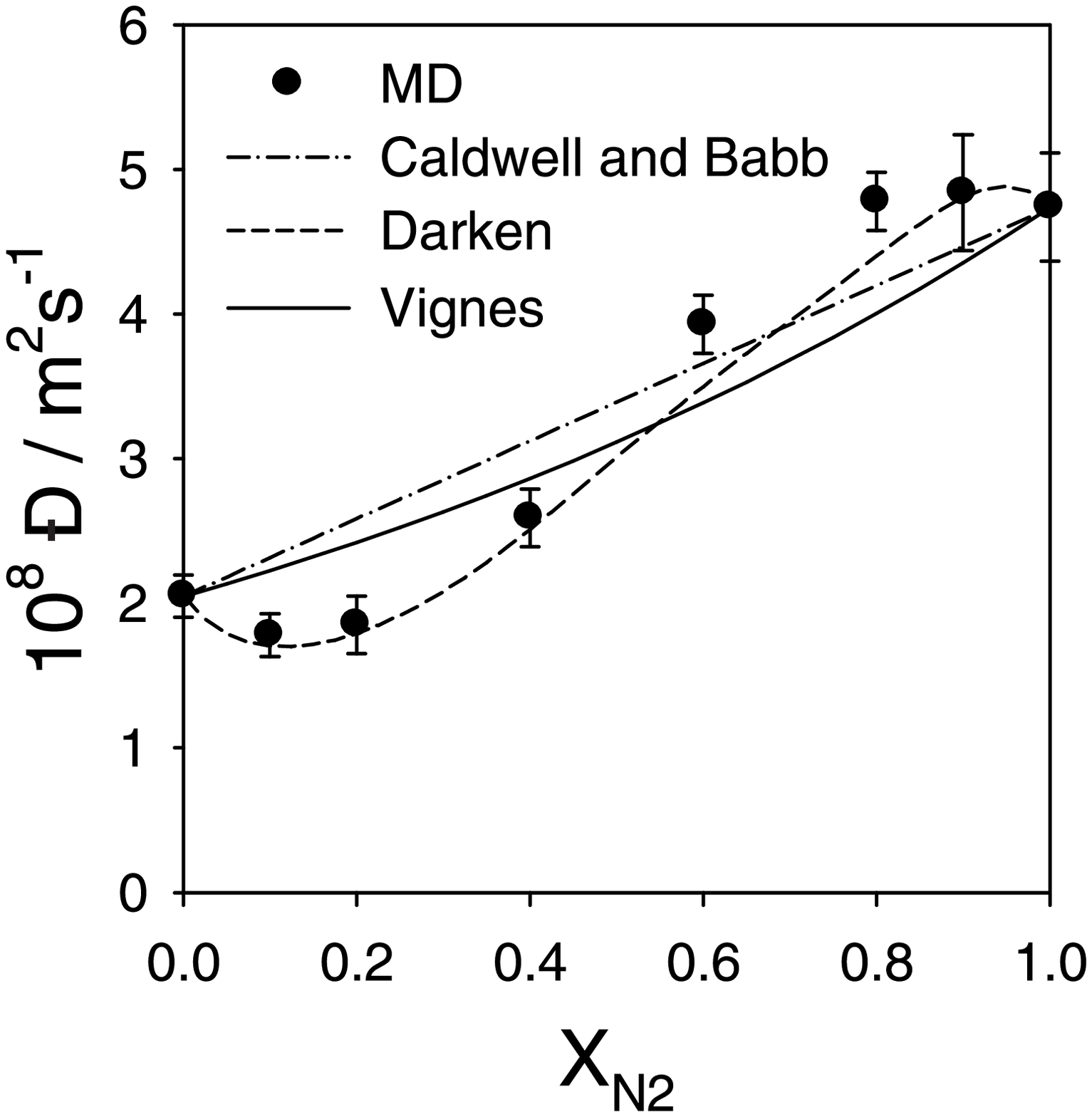}
\end{center}
\end{figure}

Figure \ref{fig4} shows the results for the binary MS diffusion
coefficients of the mixture N$_2$+C$_2$H$_6$. In this case, the MS
diffusion coefficients lie below the linear interpolation of the
infinite dilution diffusion coefficients for mole fractions smaller than
0.5 and lie above the linear interpolation for mole fractions larger than 0.5.
The results of Darken's equation agree well with the simulation data.
The average deviation is only about 6\%. The equation of Vignes
fails to reproduce the shape of the curve,
which results in an average deviation of about 20\%. The
deviations between the simulation results and the correlation of Caldwell and
Babb are also about 20\%.

\begin{figure}[ht]
\caption[Binary MS diffusion coefficients for the mixture
CO$_2$+C$_2$H$_6$ at 253.15 K and 20 MPa as predicted by molecular
simulation in comparison to empirical equations.]{}
\label{fig5}
\begin{center}
\includegraphics[width=150mm,height=200mm]{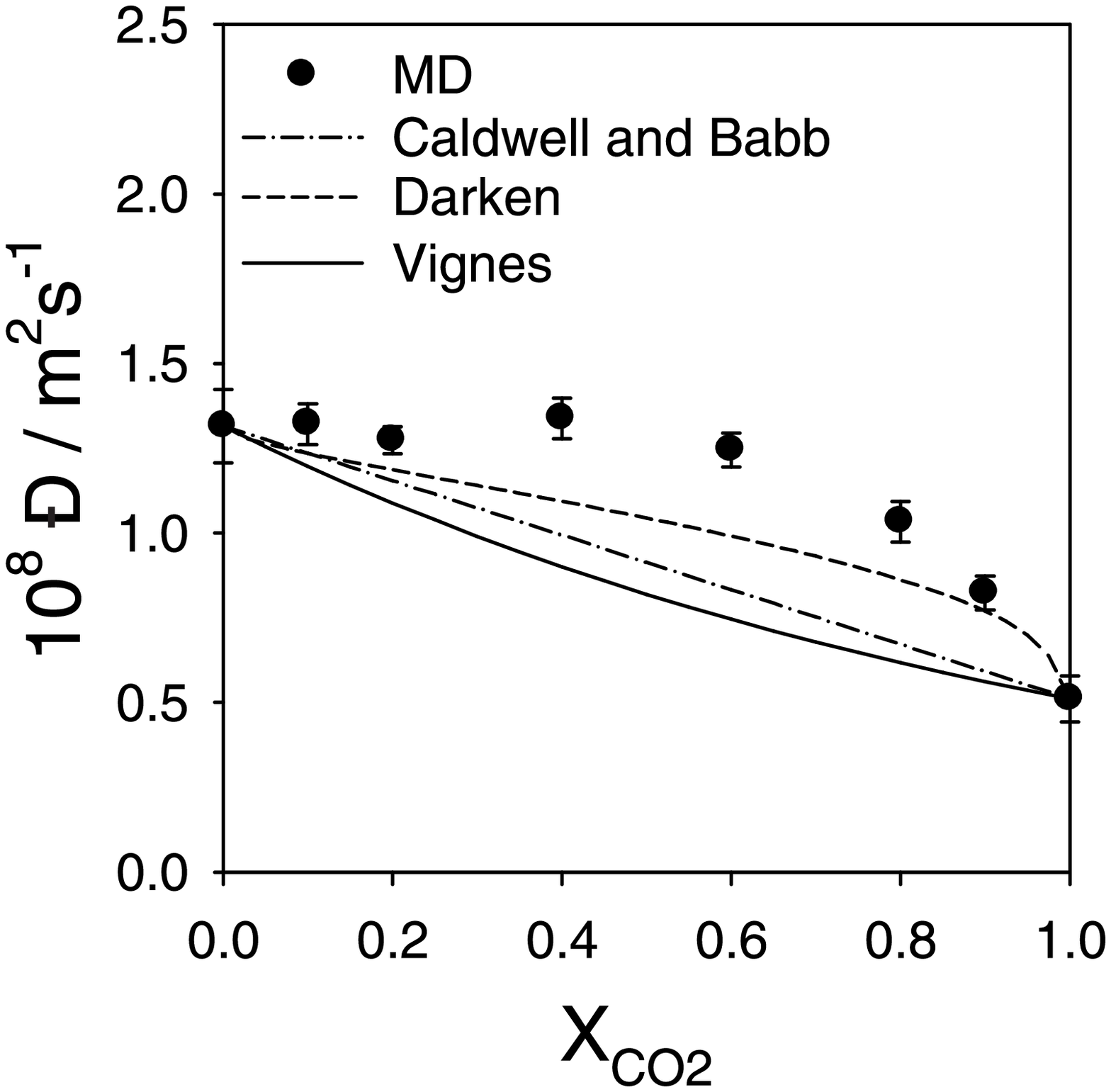}
\end{center}
\end{figure}

Figure \ref{fig5} shows the results for the binary MS diffusion
coefficients of the mixture CO$_2$+C$_2$H$_6$. In this case, the MS
diffusion coefficients lie above the linear interpolation between
the infinite dilution diffusion coefficients (Caldwell and Babb) over the
whole composition range. Also, Darken's equation here yields
the best results with an average deviation of 12\%, whereas the
equations of Caldwell and Babb and of Vignes yield deviations of 23\% and 28\%,
respectively. Again Vignes' equation does not reproduce the sign
of the deviations from the linear interpolation correctly.

Figures \ref{fig3}, \ref{fig4}, and \ref{fig5} show that the
curvature of the MS diffusion coefficient is a function of the
mole fraction, depending qualitatively on the mixture. It
can be concave, with a positive deviation from the linear course,
or convex with a negative deviation, or both. The investigated
mixtures are not strongly polar, and also in the 2CLJQ models only
quadrupolar interactions are present. However, the MS diffusion
coefficients of these mixtures can not be well represented by the
equations of Caldwell and Babb or of Vignes, that are often claimed
to be adequate for such simple mixtures
\cite{vignes}.\\
Dullien \cite{dullien} compared the predictions of Vignes'
equation with experimental data, and also found that in many
cases, where the mixtures were nonassociating, the
equation of Vignes was not able to predict the binary MS diffusion
coefficients correctly. The equation of Darken shows the best performance in
all cases. That is due to the fact that it uses more information
than the other two. Moreover, it can be shown that it is exact if
the cross correlations between different particles of the same
species and particles of different species are neglected
\cite{schoen}. Unfortunately, Darken's equation is of little
use for most practical applications.

3.3. Binary Self-diffusion Coefficients

\begin{figure}[ht]
\caption[Self- and binary MS diffusion coefficients for the
mixture N$_2$(1)+CO$_2(2)$ at 253.15 K and 20 MPa as predicted by
molecular simulation: ${\triangle}$ $D_{1}$; ${\triangledown}$
$D_{2}$; ${\bullet}$ ${\DMS_{12}}$. The lines serve as guide for the
eye.]{} \label{fig6}
\begin{center}
\includegraphics[width=150mm,height=200mm]{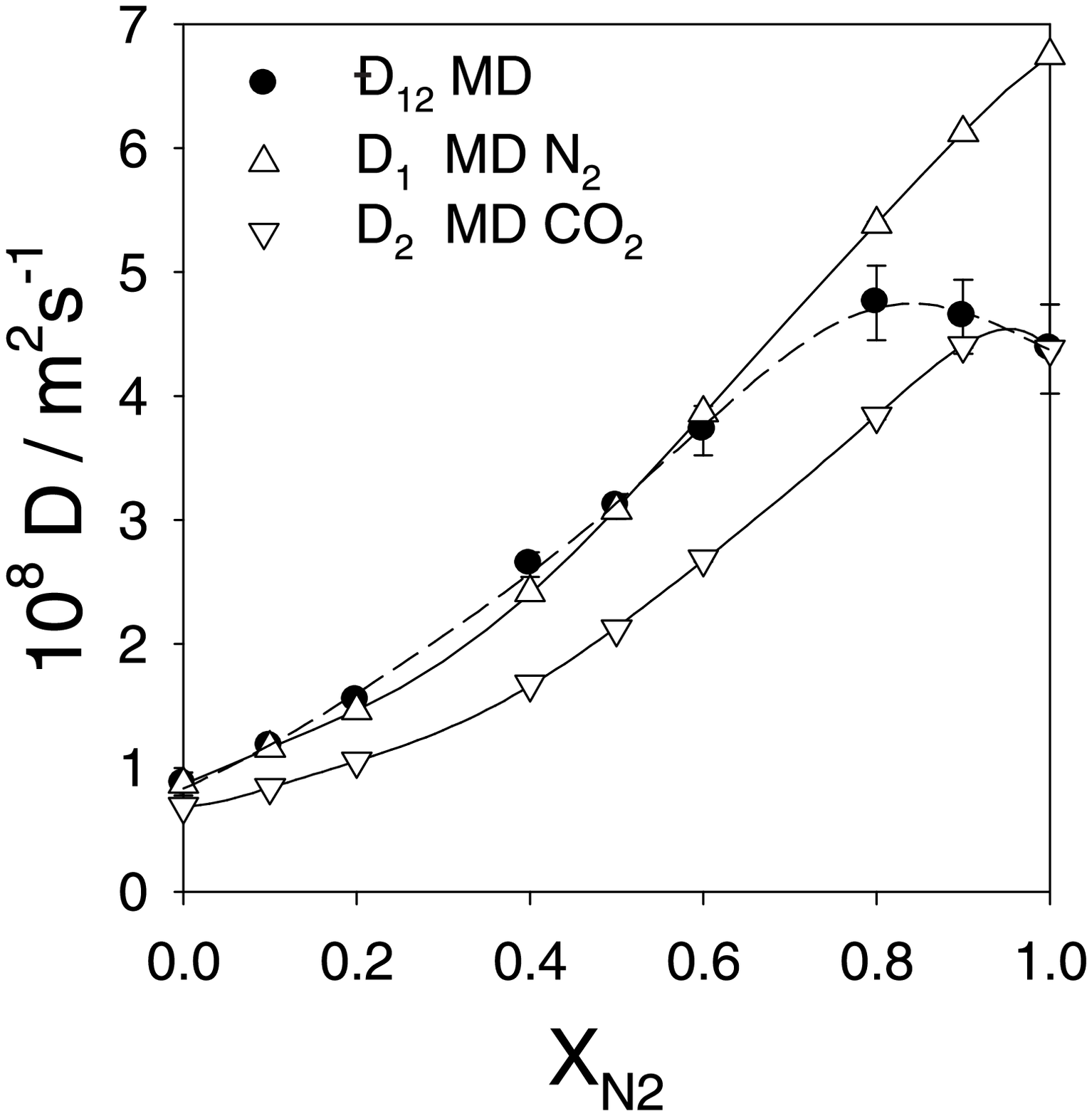}
\end{center}
\end{figure}

\begin{figure}[ht]
\caption[Self- and binary MS diffusion coefficients for the
mixture N$_2(1)$+C$_2$H$_6(2)$ at 253.15 K and 20 MPa as predicted by
molecular simulation: ${\triangle}$ $D_{1}$; ${\triangledown}$
$D_{2}$; ${\bullet}$ ${\DMS_{12}}$. The lines serve as guide for the
eye.]{} \label{fig7}
\begin{center}
\includegraphics[width=150mm,height=200mm]{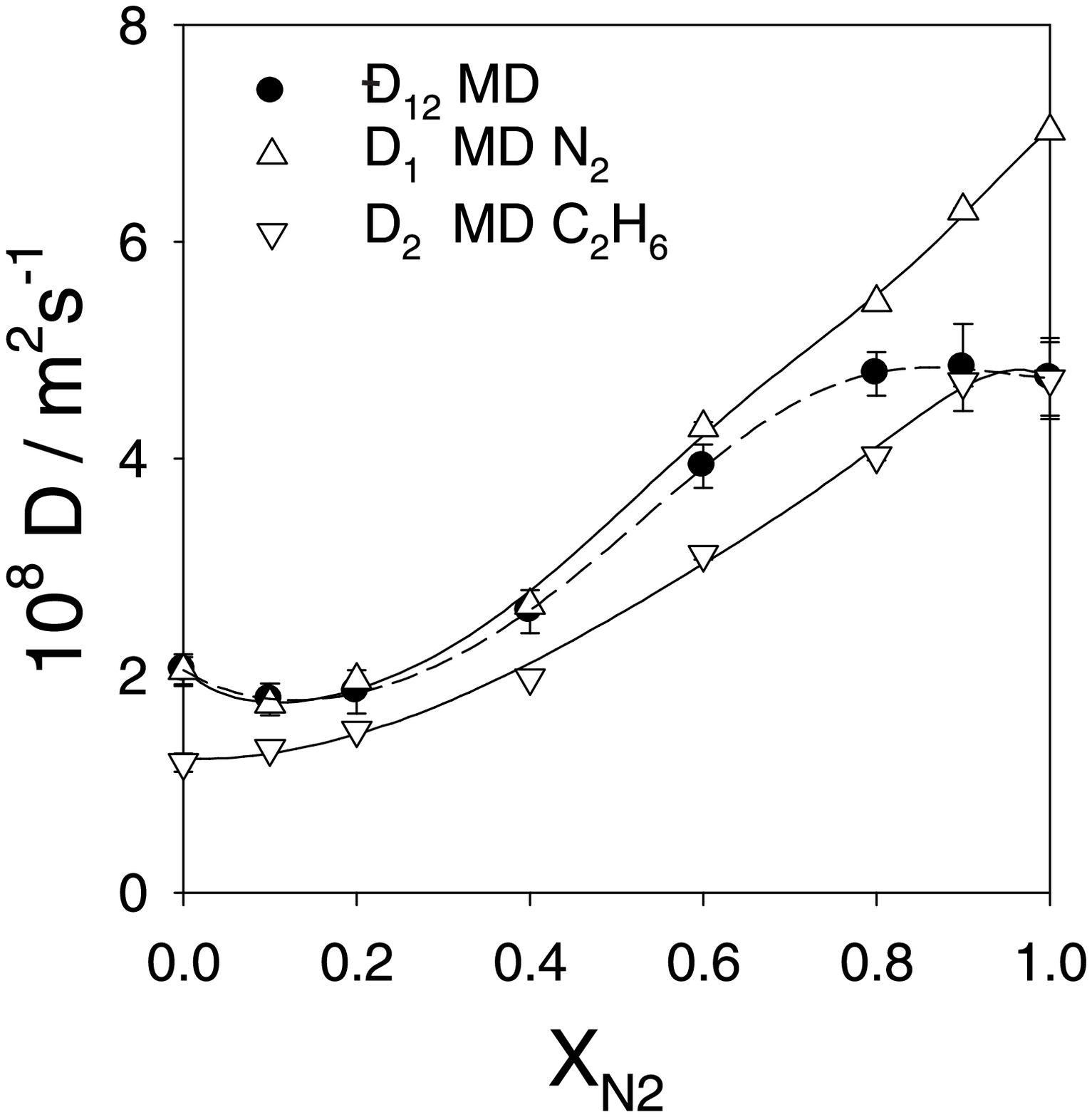}
\end{center}
\end{figure}

\begin{figure}[ht]
\caption[Self- and binary MS diffusion coefficients for the
mixture CO$_2(1)$+C$_2$H$_6(2)$ at 253.15 K and 20 MPa as predicted
by molecular simulation: ${\triangle}$ $D_{1}$; ${\triangledown}$
$D_{2}$; ${\bullet}$ ${\DMS_{12}}$. The lines serve as guide for the
eye.]{} \label{fig8}
\begin{center}
\includegraphics[width=150mm,height=200mm]{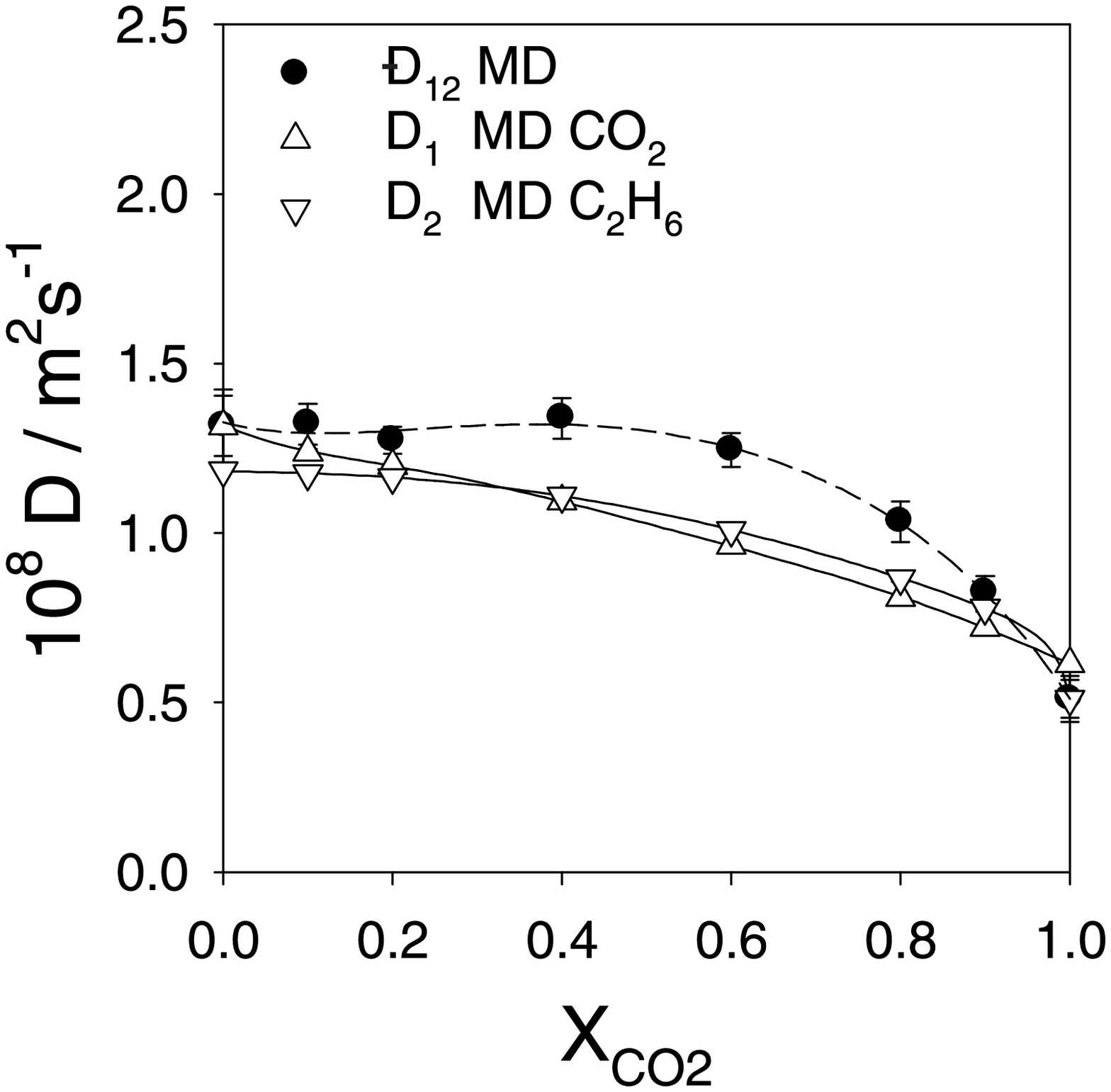}
\end{center}
\end{figure}

Figures \ref{fig6}-\ref{fig8} show the results for self-diffusion coefficients
of the pure components in the mixtures N$_2$+CO$_2$,
N$_2$+C$_2$H$_6$, and CO$_2$+C$_2$H$_6$ at 253.15 K and 20 MPa,
together with those for the binary MS diffusion coefficients.
Whereas for N$_2$+CO$_2$ and N$_2$+C$_2$H$_6$ the MS diffusion coefficients
can qualitatively be described by a simple interpolation as
indicated by Darken's equation. The situation is different for
CO$_2$+C$_2$H$_6$, cf. Fig. \ref{fig8}. The self-diffusion
coefficients are almost equal for that mixture at all
compositions. Nevertheless, the MS diffusion coefficient from
the simulations is larger so that Eq. (\ref{darken}) is inappropriate.

\clearpage
%\section{Conclusion}
\textbf{4. CONCLUSION}\\
In the present work, molecular dynamics simulation and the
Green-Kubo formalism were used to calculate self- and binary MS
diffusion coefficients for a class of fluids modeled by the 2CLJQ
intermolecular potential. The potential parameters were taken from previous work
\cite{vrabec,stoll2}
where they were adjusted to experimental vapor-liquid equilibria only.
Eight pure fluids, i.e. F$_2$, N$_2$,
CO$_2$, CS$_2$, C$_2$H$_6$, C$_2$H$_4$, C$_2$H$_2$ and three
binary mixtures, i.e. N$_2$+CO$_2$, N$_2$+C$_2$H$_6$ and
CO$_2$+C$_2$H$_6$, were studied. Self-diffusion coefficients
are reported with statistical uncertainties smaller than 1\%. These results do not consider
corrections due to the long-time tail, the error due to it
is estimated to be about 3\%. Deviations between the predicted and the experimental data do not exceed 20\%.
The correction due to the finite size of the simulated system increases the
self-diffusion coefficients typically by 10\%. With this correction an even better
agreement can be expected for most fluids. Exceptions are F$_2$, SF$_6$, and C$_2$H$_6$
for which the deviations would increase.

For the binary mixtures, predictions from the simulations are only
compared to results from the equations of Darken, Caldwell and Babb,
and Vignes, as experimental data were not available. The self-diffusion coefficients
are reported with statistical uncertainties smaller than 1\% and the binary MS diffusion
coefficients are reported with statistical uncertainties of about 10\%. In agreement with
previous findings \cite{fernandez1}, Darken's equation yields the best agreement in all
cases with average deviations of only 10\%. Unfortunately, this equation requires
self-diffusion coefficients in the mixture as input
data. The two simple equations of Caldwell and Babb and of Vignes
which use infinite dilution diffusion coefficients as input data, fail to
predict the shape of the composition dependence of the MS
diffusion coefficients, which shows a strong curvature, despite
the fairly simple molecules studied here.
This indicates that more accurate correlations for the prediction of MS diffusion
coefficients are needed. For their development, molecular simulation is a useful tool, as it can relate
molecular properties, i.e. polarity, anisotropy etc., to diffusion
coefficients.

\clearpage

\clearpage
% Tables
\begin{table}[t]
\noindent \caption{Potential Parameters for the Pure Fluids Used in This
Work$^{a,b}$.}\label{T1}
\bigskip
\begin{center}
\begin{tabular}{lccccc}
\hline Fluid &  \0\0$\sigma$ / (\r{A}) & $\epsilon$/k$_{\rm B}$ /
(K) & \textit{L} / (\r{A})& 10$^{20}$Q /(C $\cdot$ m$^2$) & $M$ / (g $\cdot$ mol$^{-1}$)
\\\hline
 $\rm F_2 $    & 2.8258 &  \052.147 &  1.4129 & 2.9754  & \038.00    \\
 $\rm N_2 $    & 3.3211 &  \034.897 &  1.0464 & 4.8024  & \028.01    \\
 $\rm CO_2 $   & 2.9847 &  133.22\0 &  2.4176 & 12.6549 & \044.01    \\
 $\rm CS_2 $   & 3.6140 &  257.68\0 &  2.6809 & 13.0081 & \076.14    \\
 $\rm C_2H_6$  & 3.4896 &  136.99\0 &  2.3762 & 2.7609  & \030.07    \\
 $\rm C_2H_4$  & 3.7607 &  \076.950 &  1.2695 & 14.4468 & \028.05    \\
 $\rm C_2H_2$  & 3.5742 &  \079.890 &  1.2998 & 16.9218 & \028.05    \\
 $\rm SF_6$    & 3.9615 &  118.98\0 &  2.6375 & 26.7074 &  146.06    \\
 \hline
\end{tabular}
\end{center}
\end{table}
\footnotetext[1]{Values taken from ref. 15.} \footnotetext[2]{The
molar mass $M$ was taken from ref. 27.} \clearpage

\begin{table}[t]
\noindent \caption{Self-diffusion Coefficients for
F$_2$, N$_2$, CO$_2$, CS$_2$, C$_2$H$_6$, C$_2$H$_4$, C$_2$H$_2$ and SF$_6$ Predicted by Molecular Simulation
Compared to Experimental Data.}\label{T2}
\bigskip
\begin{center}
\begin{tabular}{lccc|ccccc}
\hline
$\rm F_2$ &$\0\0\0$& $\0\0\0$ &
$\0\0\0$ & $\rm N_2$ & &  &  & \\
\hline ~$T$ & 10$^{-3}$ $\rho$ & 10$^{9}$ $D^{exp.}$ &
10$^{9}$ $D^{sim.}$ & $T$ & 10$^{-3}$ $\rho$ & 10$^{9}$ $D^{exp.}$ &
10$^{9}$ $D^{sim.}$ \\[-4mm]
(K) & (mol~$\cdot$~m$^{-3}$) & (m$^2$$\cdot$
s$^{-1}$) & (m$^2$$\cdot$ s$^{-1}$) & (K) & (mol~$\cdot$~m$^{-3}$) & (m$^2$$\cdot$
s$^{-1}$) & (m$^2$$\cdot$ s$^{-1}$) \\\hline
\054.0  &  44.824 & 0.569   &  0.569(2)  & 77.0  &  28.861 & 2.526 &  2.923(8) \\[-3.8mm]
\062.0  &  43.497 & 1.05\0  &  0.905(2)  & 80.0  &  28.380 & 2.996 &  3.309(5) \\[-3.8mm]
\070.0  &  42.166 & 1.69\0  &  1.361(3)  & 83.0  &  27.870 & 3.509 &  3.757(8) \\[-3.8mm]
\078.0  &  40.787 & 2.46\0  &  1.903(6)  & 85.0  &  27.526 & 3.875 &  4.03(1)\0  \\[-3.8mm]
\088.0  &  38.968 & 3.57\0  &  2.793(8)  & 88.0  &  27.006 & 4.459 &  4.63(1)\0  \\[-3.8mm]
\096.0  &  37.405 & 4.55\0  &  3.575(6)  & 90.0  &  26.643 & 4.871 &  4.93(1)\0  \\[-3.8mm]
105.0   &  35.497 & 5.73\0  &  4.74(1)\0 & 93.0  &  26.079 & 5.522 &  5.54(1)\0  \\\hline
$\rm CO_2$ &$\0\0\0$& $\0\0\0$ &
$\0\0\0$ & $\rm C_2H_4$ & &  &  & \\
\hline

273.0 &  21.102\0\0  & \013.50    & 10.41(3)\0   & 298.15 & \04.4955   & 113.6\0\0  & 110.0(3)\0    \\[-3.8mm]
273.0 &  21.453\0\0  & \013.00    & 10.39(6)\0   & 298.15 & \06.2923   & \079.91\0  & \076.1(3)\0   \\[-3.8mm]
273.0 &  22.333\0\0  & \011.70    & \09.274(3)  & 298.15 & \08.0927   & \062.10\0  & \056.6(1)\0   \\[-3.8mm]
273.0 &  23.046\0\0  & \010.70    & \08.156(4)  & 298.15 & \09.8895   & \049.45\0  & \042.1(2)\0   \\[-3.8mm]
273.0 &  23.460\0\0  & \010.00    & \07.890(2)  & 298.15 & 11.690\0   & \039.58\0  & \033.20(8)  \\[-3.8mm]
273.0 &  23.900\0\0  & \0\09.50   & \07.518(4)  & 298.15 & 13.487\0   & \031.21\0  & \025.5(1)\0   \\[-3.8mm]
 &    & &     & 298.15 & 15.283\0   & \024.08\0  & \019.90(7)  \\[-3.8mm]
 &   &    &   & 298.15 & 17.084\0   & \018.20\0  & \014.60(3)  \\[-3.8mm]
 &   &    &   & 298.15 & 18.881\0   & \013.44\0  & \010.70(3)  \\[-3.8mm]
 &   &    &   & 298.15 & 20.681\0   & \0\09.927  & \0\07.28(9)  \\\hline

$\rm C_2H_6$ &$\0\0\0$& $\0\0\0$ &
$\0\0\0$ & $\rm CS_2$ & &  &  & \\\hline

273.0 &  15.431 & 14.6\0   &   14.40(4)\0   &   298.2 &  16.489 & 4.26  &  3.209(7) \\[-3.8mm]
273.0 &  16.550 & 11.8\0   &   11.71(6)\0   &   298.2 &  17.019 & 3.64  &  2.653(8) \\[-3.8mm]
273.0 &  17.968 & \08.91 &   \09.008(4)  &   298.2 &  17.514 & 3.21  &  2.264(7) \\[-3.8mm]
273.0 &  18.902 & \07.24 &   \07.230(2)  &   298.2 &  18.031 & 2.61  &  1.867(5) \\[-3.8mm]
273.0 &  19.609 & \06.27 &   \05.870(1)  &   298.2 &  18.543 & 2.23  &  1.532(5) \\\hline
$\rm C_2H_2$ &$\0\0\0$& $\0\0\0$ &
$\0\0\0$ & $\rm SF_6$ & &  &  & \\\hline

192.0 &  23.754 & \03.74 &   \02.91(1)  &   240.0  & 12.091  & \03.35  &  \03.52(3) \\[-3.8mm]
197.0 &  23.463 & \04.26 &   \03.37(1)  &   250.0  & 11.653  & \03.94  &  \04.28(2)  \\[-3.8mm]
202.0 &  23.167 & \04.82 &   \03.95(1)  &   260.0  & 11.221  & \04.66  &  \04.89(4)  \\[-3.8mm]
207.0 &  22.863 & \05.43 &   \04.35(1)  &   270.0  & 10.742  & \05.59  &  \06.03(2)  \\[-3.8mm]
212.0 &  22.554 & \06.07 &   \04.66(1)  &   280.0  & 10.201  & \06.71  &  \07.49(4)  \\[-3.8mm]
217.0 &  22.237 & \06.76 &   \05.44(1)  &   290.0  & \09.606 & \08.29  &  \08.87(5)  \\[-3.8mm]
222.0 &  21.912 & \07.49 &   \05.95(2)  &   300.0  & \08.846 & 10.5\0    &  11.00(2)  \\[-3.8mm]
310.0 & \07.826 & 14.4\0 &   14.50(5)   &          &         &         &           \\\hline

\end{tabular}
\end{center}
\end{table}

\begin{table}[t]
\noindent \caption{Self-diffusion and Binary MS Diffusion Coefficients of
the Binary Mixtures N$_2$+CO$_2$, N$_2$+C$_2$H$_6$, and
CO$_2$+C$_2$H$_6$ at 253.15 K and 20 MPa Predicted by Molecular
Simulation.}\label{T3}
\bigskip
\begin{center}
\begin{tabular}{lccccc}
\hline $\0\0\0\0\0\0$ & $\0\0\0\0\0\0$ & $\0\0\0\0\0$ $\rm
N_2(1)+CO_2(2)$ $\0\0\0\0\0$ &
$\0\0\0\0\0\0$ & $\0\0\0\0\0\0$ \\
\hline $x_1$ & 10$^{-3}$ $\rho$ & 10$^{9}$ $D_{1}$ &
10$^{9}$ $D_{2}$ & 10$^{9}$ $\DMS_{12}$ \\[-4mm]
 & (mol~$\cdot$~m$^{-3}$) & (m$^2$$\cdot$ s$^{-1}$) & (m$^2$$\cdot$
s$^{-1}$) & (m$^2$$\cdot$ s$^{-1}$) \\\hline
0.0 & 24.08 &  \08.7(9)\0\0   & \06.86(2)\0  & \08.7(9) \\[-3.5mm]
0.1 & 22.96 &   11.57(7)\0    & \08.40(3)\0  &  11.7(5) \\[-3.5mm]
0.2 & 21.47 &   14.60(7)\0    &  10.55(4)\0  &  15.4(3) \\[-3.5mm]
0.4 & 17.73 &   24.13(10)     &  16.74(2)\0  &  26(1)\0 \\[-3.5mm]
0.5 & 15.59 &   30.77(8)\0    &  21.19(7)\0  &  31(1)\0 \\[-3.5mm]
0.6 & 13.60 &   38.63(6)\0    &  26.83(19)  &  37(2)\0 \\[-3.5mm]
0.8 & 10.90 &   53.87(20)     &  38.39(25)  &  47(3)\0 \\[-3.5mm]
0.9 & 9.968 &   61.25(20)     &  44.08(24)  &  46(3)\0 \\[-3.5mm]
1.0 & 9.356 &   67.49(13)     &  43(4)\0\0\0 &  43(4)\0 \\\hline
$\0\0\0\0\0\0$ & $\0\0\0\0\0\0$ & $\0\0\0\0\0$ $\rm
N_2(1)+C_2H_6(2)$ $\0\0\0\0\0$ &
$\0\0\0\0\0\0$ & $\0\0\0\0\0\0$ \\\hline
0.0 & 13.87\0 &  20(1)\0\0\0    & 11.97(4)\0  &  20(1) \\[-3.5mm]
0.1 & 13.85\0 &  17.39(9)\0 & 13.27(2)\0  &  18(1) \\[-3.5mm]
0.2 & 13.74\0 &  19.58(16)  & 14.98(4)\0  &  18(1) \\[-3.5mm]
0.4 & 12.91\0 &  26.48(6)\0 & 19.81(11)   &  26(2) \\[-3.5mm]
0.6 & 11.21\0 &  42.88(50)  & 31.18(20)   &  39(3) \\[-3.5mm]
0.8 & \09.302 &  54.42(11)  & 40.25(38)   &  48(4) \\[-3.5mm]
0.9 & \08.566 &  62.88(17)  & 47.08(40)   &  48(4) \\[-3.5mm]
1.0 & \08.059 &  70.21(20)  & 47(4)\0\0\0 &  47(4) \\\hline $\0\0\0\0\0\0$ &
$\0\0\0\0\0\0$ & $\0\0\0\0\0 $ $\rm CO_2(1)+C_2H_6(2)$ $\0\0\0\0\0$
&
$\0\0\0\0\0\0$ & $\0\0\0\0\0\0$ \\
\hline
0.0  & 16.10 & 13(1)\0\0\0 & 11.84(1)\0  & 13(1)\0  \\[-3.5mm]
0.1  & 16.47 & 12.38(14)   & 11.75(6)\0  & 13(1)\0  \\[-3.5mm]
0.2  & 16.88 & 11.98(5)\0    & 11.64(6)\0  & 13(1)\0  \\[-3.5mm]
0.4  & 17.95 & 10.93(6)\0    & 11.09(8)\0  & 13(1)\0  \\[-3.5mm]
0.6  & 19.49 & \09.63(4)\0   & 10.08(6)\0  & 12(1)\0  \\[-3.5mm]
0.8  & 21.59 & \08.09(3)\0   & \08.66(9)\0 & 10.3(8)  \\[-3.5mm]
0.9  & 22.94 & \07.21(1)\0   & \07.78(14)& \08.2(7) \\[-3.5mm]
1.0  & 24.66 & \06.15(6)\0   & \05.1(7)\0\0  & \05.1(7) \\\hline
\end{tabular}
\end{center}
\end{table}

\clearpage
% List of figures
\listoffigures

\textbf{REFERENCES}\\
$\0$1. B.J. Alder and T.E. Wainwright, {\it Phys. Rev. A} {\bf 18}:988 (1967).

$\0$2. B.J. Alder and T.E. Wainwright, {\it Phys. Rev. A} {\bf 1}:18 (1970).

$\0$3. G. Jaccuci and I.R. Mcdonald, {\it Physica A} {\bf 80}:607 (1975).

$\0$4. D. Jolly and R. Bearman, {\it Mol. Phys.} {\bf 41}:137 (1980).

$\0$5. M. Schoen and C. Hoheisel, {\it Mol. Phys.} {\bf 52}:33 (1984).

$\0$6. J.M. Stoker and R.L. Rowley, {\it J. Chem. Phys.} {\bf 91}:3670 (1989).

$\0$7. R.L. Rowley and J.M. Stoker, {\it Int. J. Thermophys.} {\bf 12}:501 (1991).

$\0$8. G.A. Fern\'{a}ndez, J. Vrabec, and H. Hasse, {\it Int. J. Thermophys.} {\bf25}:175 (2004).

$\0$9. G.A. Fern\'{a}ndez, J. Vrabec, and H. Hasse, {\it Fluid Phase Equilib.} {\bf 221}:157 (2004).

10. I.R. McDonald and K. Singer, {\it Mol. Phys.} {\bf 23}:29 (1972).

11. P.S.Y. Cheung and J.G. Powles, {\it Mol. Phys.} {\bf 30}:921 (1975).

12. R. Vogelsang and C. Hoheisel, {\it Phys. Chem. Liq.} {\bf 16}:189 (1987).

13. C. Hoheisel, {\it Mol. Phys.} {\bf 62}:239 (1987).

14. D. M\"{o}ller and J. Fischer, {\it Fluid Phase Equilib.} {\bf 100}:35 (1994).

15. J. Vrabec, J. Stoll, and H. Hasse, {\it J. Phys. Chem. B} {\bf 105}:12126 (2001).

16. C.S. Caldwell and A.L. Babb, {\it J. Phys. Chem.} {\bf 60}:51 (1956).

17. L.S. Darken, {\it Trans. Am. Inst. Mining Metall. Eng.} {\bf 175}:184 (1948).

18. A. Vignes, {\it Ind. Eng. Chem. Fundam.} {\bf 5}:189 (1966).

19. C.G. Gray, and K.E. Gubbins, {\it Theory of Molecular Fluids, Vol. 1: Fundamentals}, $\0\0\0$(Clarendon
Press, Oxford, 1984).

20. J. Vrabec and J. Fischer, {\it AIChE J.} {\bf 43}:212 (1996).

21. J. Vrabec, J. Stoll, and H. Hasse, {\it Mol. Sim.} {\bf 31}:215 (2005).

22. J. Stoll, J. Vrabec, and H. Hasse, {\it AIChE J.} {\bf 49}:2187 (2003).

23. M.S. Green, {\it J. Chem. Phys.} {\bf 22}:398 (1954).

24. R. Kubo, {\it J. Phys. Soc. Japan} {\bf 12}:570 (1957).

25. R. Zwanzig, {\it Ann. Rev. Phys. Chem.} {\bf 16}:67 (1965).

26. C. Hoheisel, {\it Phys. Rep.} {\bf 245}:111 (1994).

27. NIST Chemistry WebBook, http://webbook.nist.gov/chemistry.

28. R. Lustig, {\it Mol. Phys.} {\bf 65}:175 (1988).

29. J.M. Haile, {\it Molecular Dynamics Simulation} (John Wiley \&
Sons Inc., New York, $\0\0\0$1997).

30. S. Nos\'{e}, {\it Mol. Phys.} {\bf 100}:191 (2002).

31. D. Frenkel and R. Smit, {\it Understanding Molecular
Simulation} (Academic Press, San $\0\0\0$Diego, 1996).

32. H.C. Andersen, {\it J. Chem. Phys.} {\bf 72}:2384 (1980).

33. D. Fincham, N. Quirke, and D.J. Tildesley, {\it J. Chem. Phys.} {\bf 84}:4535 (1986).

34. H. Liu, C.M. Silva, and E.A. Macedo, {\it Chem. Eng. Sci.} {\bf 53}:2403 (1998).

35. D.E. O'Reilly, E.M. Peterson, D.I. Hogenboom, and C.E. Scheie, {\it J. Chem. Phys.} $\0\0\0${\bf 54}:4194
(1971).

36. K. Krynicki, E.J., Rahkamaa, and J.P. Powles, {\it Mol. Phys.} {\bf 28}:853 (1974).

37. P.E. Etesse, J.A. Zega, and R. Kobayashi, {\it J. Chem. Phys.} {\bf 97}:2022 (1992).

38. L.A. Woolf, {\it J. Chem. Soc. Faraday Trans. 1} {\bf 78}:583 (1982).

\clearpage

39. N.B. Vargaftik, Y.K. Vinogradov, and V.S. Yargin, {\it Handbook of Physical Properties $\0\0\0$ of Liquids
and Gases} (Begell House, New York, 1996).

40. E.W. Lemmon, M.O. McLinden, and M.L. Huber, {\it REFPROP} (NIST Standard $\0\0\0$ $\0\0\0$Reference Database
23, Version 7.0, 2002).

41. A. Greiner-Schmid, S. Wappmann, M. Has, and H.D. L\"{u}demann, {\it J. Chem. Phys.} $\0\0\0${\bf 94}:5643
(1991).

42. B. Arends, K.O. Prins, and N.J. Trappeniers, {\it Physica A} {\bf 107A}:307 (1981).

43. C.E. Scheie, E.M. Peterson, and D.E. O'Reilly, {\it J. Chem. Phys.} {\bf 59}:2303 (1973).

44. J.K. Tison and E.R. Hunt, {\it J. Chem. Phys.} {\bf 54}:1526 (1971).

45. T.E. Daubert and R.P. Danner, {\it Data compilation tables of properties of pure}\\
$\0\0\0${\it substances}, AIChE (American Chemical Society, New York, 1985).

46. Yeh, I. and G. Hammer, {\it J. Chem. Phys.} {\bf 108}:158 (2004).

47. F.A.L. Dullien, {\it Ind. Eng. Chem. Fundam.} {\bf 10}:41 (1971).

% Bibliography


\begin{thebibliography}{99}

\bibitem{alder1} B.J. Alder and T.E. Wainwright, {\it Phys. Rev. A.} {\bf 18}:988 (1967).

\bibitem{alder2} B.J. Alder and T.E. Wainwright, {\it Phys. Rev. A.} {\bf 1}:18 (1970).

\bibitem{jaccuci} G. Jaccuci and I.R. Mcdonald, {\it Physica A} {\bf 80}:607 (1975).

\bibitem{jolly} D. Jolly and R. Bearman, {\it Mol. Phys.} {\bf 41}:137 (1980).

\bibitem{schoen} M. Schoen and C. Hoheisel, {\it Mol. Phys.} {\bf 52}:33 (1984).

\bibitem{stoker}J.M. Stoker and R.L. Rowley, {\it J. Chem. Phys.} {\bf 91}:3670(1989).

\bibitem{rowley1}R.L. Rowley and J.M. Stoker, {\it Int. J.
Thermopys.}{\bf 12}:501(1991).

\bibitem{fernandez1}G.A. Fern\'{a}ndez, J. Varbec, and H. Hasse {\it Int. J.
Thermopys.}{\bf 25}:175(2004).

\bibitem{fernandez2}G.A. Fern\'{a}ndez, J. Varbec, and H. Hasse {\it Fluid Phase
Equilib.}{\bf 221}:157(2004).

\bibitem{mcdonald}I.R. McDonald, K. Singer, {\it Mol. Phys.} {\bf 23}:29(1972).

\bibitem{cheung}P.S.Y. Cheung, and J.G. Powles, {\it Mol. Phys.} {\bf 30}:921(1975).

\bibitem{vogelsang}R. Vogelsang, and C. Hoheisel, {\it Phys. Chem. Liq} {\bf 16}:189-203(1987).

\bibitem{hoheisel}C. Hoheisel, {\it Mol. Phys.} {\bf 62}:239(1987).

\bibitem{moeller}D. M\"{o}ller and J. Fischer, {\it Fluid Phase Equilib.} {\bf 100}:35(1994).

\bibitem{vrabec}J. Vrabec, J. Stoll and H. Hasse, {\it J. Phys. Chem. B} {\bf 105}:12126(2001).

\bibitem{caldwellbabb}C.S. Caldwell and A.L. Babb, {\it J. Phys. Chem.} {\bf 60}:51(1956).

\bibitem{darken}L.S. Darken, {\it AIME} {\bf 175}:184 (1948).

\bibitem{vignes}A. Vignes, {\it Ind. Eng. Chem. Fundam.} {\bf 2}:189(1966).

\bibitem{graygubb}C.G. Gray, and K.E. Gubbins,{\it Theory of Molecular Fluids, Vol. 1:
Fundamentals}, (Clarendon Press, Oxford, 1984).

\bibitem{vrabec2}J. Vrabec, J. Stoll, and H. Hasse, {\it J. Phys. Chem. B} submitted.

\bibitem{vrabec3}J. Vrabec and J. Fischer, {\it AiChe J.} {\bf 43}:212(1996).

\bibitem{stoll2}J. Stoll, J. Vrabec, and H. Hasse, {\it AiChe J.} {\bf 49}:2187(2003).

\bibitem{green} M.S. Green, {\it J. Chem. Phys.} {\bf 22}:398(1954).

\bibitem{kubo} R. Kubo, {\it J. Phys. Soc. Japan} {\bf 12}:570(1957).

\bibitem{zwanzig} R. Zwanzig, {\it Ann. Rev. Phys. Chem.} {\bf 16}:67(1965).

\bibitem{hoheisel1}C. Hoheisel, {\it Phys. Rep.} {\bf 245}:111(1994).

\bibitem{nist}NIST Chemistry WebBook, http://webbook.nist.gov/chemistry.

\bibitem{lustig}R. Lustig, {\it Mol. Phys.} {\bf 65}:175(1988).

\bibitem{haile}J.M. Haile, {\it Molecular Dynamics Simulation}(John Wiley \& Sons Inc.,
New York, 1997).

\bibitem{nose}S. Nos\'{e}, {\it Mol. Phys.} {\bf 100}:191(2002).

\bibitem{frenkel} D. Frenkel and R. Smit, {\it Understanding Molecular Simulation}
(Academic Press, San Diego, 1996).

\bibitem{andersen} H.C. Andersen, {\it J. Chem. Phys.} {\bf 72}:2384(1980).

\bibitem{fincham}D. Fincham, N. Quirke, and D.J. Tildesley, {\it J. Chem. Phys.}
{\bf 84}:4535(1986).

\bibitem{liu} H. Liu, C.M. Silva, and E.A. Macedo, {\it Chem. Eng. Sci.} {\bf 53}:2403 (1998).

\bibitem{oreilly}D.E. O'Reilly, E.M. Peterson, D.I. Hogenboom, and C.E. Scheie, {\it
J. Chem. Phys.} {\bf 54}:4194(1971).

\bibitem{krynicki}K. Krynicki, E.J., Rahkamaa, and J.P. Powles, {\it Mol. Phys.}
 {\bf 28}:853(1974).

\bibitem{etesse}P.E. Etesse, J.A. Zega, and R. Kobayashi, {\it J. Chem. Phys.}
{\bf 97}:2022(1992).

\bibitem{woolf}L.A. Woolf, {\it J. Chem. Soc. Faraday. Trans. 1} {\bf 78}:583(1982).

\bibitem{vargaftik}N.B. Vargaftik, Y.K. Vinogradov, and V.S. Yargin, {\it Handbook of Physical
Properties of Liquids and Gases}(Begell house, inc., New York, 1996).

\bibitem{refprop}E.W. Lemmon, M.O. McLinden, and M.L. Huber, {\it REFPROP}(NIST
Standard reference Database 23, Version 7.0, 2002).

\bibitem{greiner}A. Greiner-Schmid, S. Wappmann, M. Has, and H.D. L\"{u}demann,
{\it J. Chem. Phys.} {\bf 94}:5643(1991).

\bibitem{arends}B. Arends, K.O. Prins, and N.J. Trappeniers, {\it Physica A}
{\bf 107A}:307(1981).

\bibitem{scheie}C.E. Scheie, E.M. Peterson, and D.E. O'Reilly, {\it
J. Chem. Phys.} {\bf 59}:2303(1973).

\bibitem{tison}J.K. Tison, and E.R. Hunt, {\it J. Chem. Phys.} {\bf 54}:1526(1971).

\bibitem{daubert}T.E. Daubert,and R.P. Daner, {\it Data compilation tables of properties of pure
subtances}, AIChE,(American Chemical Society, New York, 1985).

\bibitem{inchul} Yeh, I. and G. Hammer, {\it J. Chem. Phys.} {\bf 108}:158(2004).

\bibitem{dullien} F.A.L. Dullien, {\it Ind. Eng. Chem. Fundam.} {\bf 10}:41(1971).

\end{thebibliography}
\end{document}